\documentclass[10pt,conference]{IEEEtran}

\usepackage[utf8]{inputenc}
\usepackage[T1]{fontenc}

\usepackage{amsmath}
\usepackage{amssymb}
\usepackage{graphicx}

\begin{document}

\title{Developing Programming Assignments for Teaching Quantum Computing and Quantum Programming}
\author{
  \IEEEauthorblockN{Mariia Mykhailova}
  \IEEEauthorblockA{mamykhai@microsoft.com \\ Microsoft Quantum, United States}
}

\IEEEoverridecommandlockouts \IEEEpubid{\begin{minipage}{\textwidth}\ \\[12pt]\ \copyright2022 IEEE. Personal use of this material is permitted. Permission from IEEE must be obtained for all other uses, in any current or future media, including reprinting/republishing this material for advertising or promotional purposes, creating new collective works, for resale or redistribution to servers or lists, or reuse of any copyrighted component of this work in other works.\hfill \end{minipage}}

\maketitle

\begin{abstract}
    
This report describes a variety of programming assignments that can be used to teach quantum computing in a practical manner. These assignments let the learners get hands-on experience with all stages of quantum software development process, from solving quantum computing problems and implementing the solutions to debugging the programs, performing resource estimation, and running the code on quantum hardware.
\end{abstract}

\section{Introduction}

Quantum computing is an emerging computing paradigm that relies on quantum-mechanical phenomena such as superposition and entanglement to perform computations.
The rapid rise of interest and investments in quantum information science and engineering (QISE) has led to increasing demand in a quantum-trained workforce.

Recent assessments of the needs of quantum industry\cite{QEDC-workforce-assessment}\cite{Aiello_2021} identified quantum software engineering and application development as one of the essential skills required for certain types of industry jobs.
Surveys of the university programs offering Master-level education in QISE\cite{Aiello_2021}\cite{Frantz_2020} show that quantum programming is frequently included as either a standalone course or one of the topics in an introductory course.

In this report I describe a variety of programming and software exploration assignments that can be used to emphasize hands-on learning and software development in a quantum computing course, based on my experience developing programming assignments for two courses: 

\begin{itemize}
    \item an undergraduate course "CSE490Q: Introduction to Quantum Computing and Quantum Programming in Q\#" delivered by Krysta Svore at the University of Washington in 2019\cite{UW}, and
    \item a graduate course "CSYE6305: Introduction to Quantum Computing with Applications" delivered by me at the Northeastern University in 2020.
\end{itemize}

Both courses focused on teaching quantum computing through a software engineering lens rather than a physics one. They targeted computer science students with previous exposure to linear algebra and programming but not physics.

In both courses student performance was evaluated using programming assignments and open-ended final projects instead of written exams\cite{UW}. In this report I focus on the different types of programming assignments that were offered in these courses in the past or can be offered in the future. Final projects are a concept familiar from computer science education, and emphasize students' initiative and creativity, so they are of less interest in the context of this report.

The assignments described in this report were developed using Q\#\cite{rwdsl2018}, a high-level domain-specific programming language designed for expressing quantum algorithms, and the Microsoft Quantum Development Kit (QDK)\footnote{https://docs.microsoft.com/azure/quantum/}, an open-source software development kit that includes a Q\# compiler, a variety of specialized quantum simulators and resource estimators, and other development tools essential for quantum software design and development.
This kit allows the user to run quantum programs on quantum hardware and cloud simulators from multiple providers via Azure Quantum\footnote{https://azure.com/quantum/}, enabling a broader spectrum of programming assignments.

I hope that this work will inspire and enable more instructors to adopt a similar software-driven approach to delivering quantum computing courses.

\IEEEpubidadjcol

\section{Quantum programming problems}

The most broadly used type of programming assignments requires the students to apply the theoretical concepts and algorithms they have learned to solving practical programming problems.
In our courses, such programming assignments were implemented in a way that supports automatic grading. 

Each task describes a concrete quantum computing problem and provides the signature of a Q\# operation that matches the description of the given problem. 
The student has to solve the problem and replace the "\texttt{// ...}" comment in the body of the operation with Q\# code that implements the solution.

A large collection of problems and solutions of this type is available as the Quantum Katas\footnote{https://github.com/Microsoft/QuantumKatas/} – an open-source project that offers tutorials and exercises on a variety of introductory topics in quantum computing.
It was used in both our courses, as well as in multiple courses at other universities\cite{Sotelo}.

\subsection{Automatic grading}

Once the student turns in the file with the Q\# solutions, the assignment can be graded automatically using a predefined testing harness - a Q\# project that runs the solutions on a certain set of tests and validates that their results match the expected ones. 

The testing harnesses rely on the quantum simulators and program testing tools available in the Microsoft Quantum Development Kit\cite{MS21}.
The solutions are evaluated based on whether their effects match the tasks' requirements, rather than on the exact paths they take to accomplish the tasks.

Microsoft Quantum Development Kit offers testing tools that allow the instructor to implement multiple kinds of program validations easily\cite{MS21}: 

\begin{itemize}
    \item check that the quantum state of the program matches the expected state,
    \item check that the operation implements the expected unitary,
    \item check that the operation implements the quantum oracle that matches the expected classical function,
    \item check that the amount of resources consumed by the program, such as the maximum number of qubits allocated or the number of oracle calls used, is within the given limits, and many others.
\end{itemize}

Note that due to the probabilistic nature of a lot of quantum algorithms creating automated testing harnesses for some kinds of tasks can be challenging. This can be addressed by offering individual building blocks of the algorithm as standalone tasks in the assignment and validating them separately (for example, validating the implementation of an oracle rather than that of an end-to-end Grover's search algorithm), or by running statistical tests on the algorithm results.

This approach to programming assignments allows the instructors to automate the grading process, significantly reducing the workload on the instructors and the teaching assistants.
Additionally, the testing harnesses for the assignments can be shared with the students as part of the assignment, providing them immediate feedback on their solutions.

\subsection{Quantum programming task example}

\begin{figure}
    \centering
    \includegraphics[width=.95\linewidth]{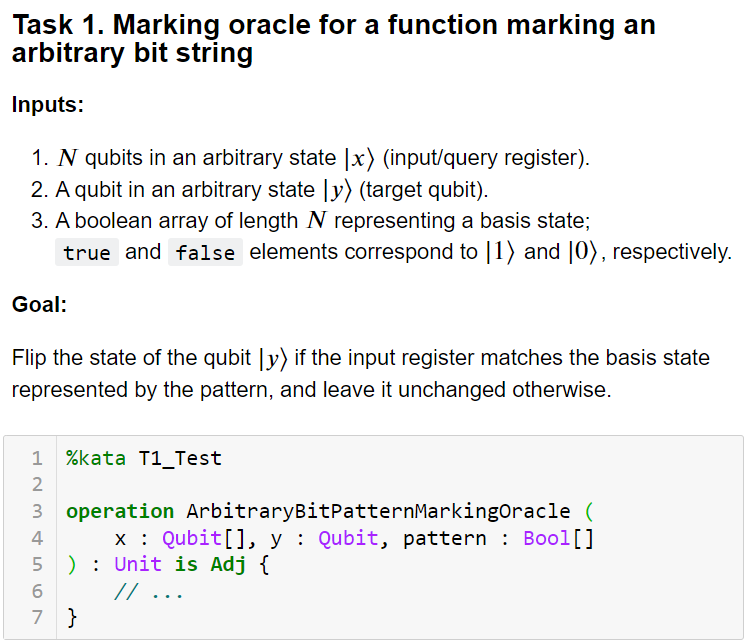}
    \caption{Marking oracle implementation task that asks the student to replace the "\texttt{// ...}" comment with Q\# code that implements the described oracle}
    \label{fig:marking-oracle-task}
\end{figure}

Figure~\ref{fig:marking-oracle-task} shows an example of a task that asks the student to implement a marking oracle for the function that marks the given bit string, that is, $f(x) = 1$ if the bit representation of $x$ matches the given pattern, and $0$ otherwise. 
The corresponding Q\# operation takes three input parameters - the qubit array representing the input of the oracle, the qubit representing the output of the oracle, and the Boolean array representing the pattern to be marked by the oracle - and does not produce an output. Since it should implement a unitary transformation, it acts by changing the state of the qubits passed to it as the arguments rather than by producing an output.

The testing harness for this task can use the full state simulator or the Toffoli simulator\footnote{https://docs.microsoft.com/azure/quantum/machines/}. 
It can run the solution on each of the basis states in turn and to check that the state of the input and output qubits after the oracle application is as expected, that is, the state of the input qubits did not change and the state of the output qubit matches the value of the function on the input encoded in the basis state. 
As long as the testing harness also checks that the solution does not use measurements, we can use the linearity of quantum operations to conclude that correct behavior of the oracle on all basis state inputs means that it will also behave correctly on all superposition inputs, and thus the solution is correct. 
At the same time, the student has the liberty to choose any solution that implements the correct unitary transformation.

\section{Debugging quantum programs}

Assignments that focus on debugging quantum programs aim to deepen the students' proficiency with the programming tools used in the course and at the same time to introduce them to this essential step of the quantum software development cycle.

The errors found in quantum programs can be grouped in three types depending on the ways they manifest:

\begin{itemize}
    \item Syntax errors: errors in the syntax of the program that prevent the code from being compiled and executed. For compiled languages such as Q\#, syntax errors can be detected at compile-time or, if the IDE used to work with the code supports IntelliSense, at development time.
    \item Runtime errors: errors in the logic of the program that are not detected at compile-time and manifest as failures during program execution. Examples of runtime errors include classical errors, such as accessing array elements using an invalid index or attempting to divide by zero, and quantum-specific errors, such as using the same qubit as both the control and the target for a controlled gate. Identifying runtime errors requires fixing all syntax errors, running the program to get the error message, and tracking it back to its source.
    \item Logical errors: errors in the logic of the program that do not manifest as failures during either compile-time or runtime, but rather produce incorrect results. Identifying logical errors requires fixing all syntax and runtime errors, running the program to get its results, analyzing these results to check their correctness, and identifying the root causes of incorrect results. For probabilistic algorithms or errors that show up only sometimes (for example, using a wrong basis to perform the measurements to get the final result of the algorithm will sometimes yield correct results), identifying the errors can involve rather sophisticated statistical analysis.
\end{itemize}

Such assignments can be useful after the first few weeks of the course, once the students move on from the basic concepts of quantum computing to simple algorithms. 
They allow the instructor to bring attention to frequent mistakes done by the students in their homework (for example, missing \texttt{open} statements or confusion about the syntax of the \texttt{Controlled} functor), to help them learn to debug their future code, and of course to check their understanding of the algorithms and the practical aspects of their applications.

\begin{figure}
    \centering
    \includegraphics[width=.95\linewidth]{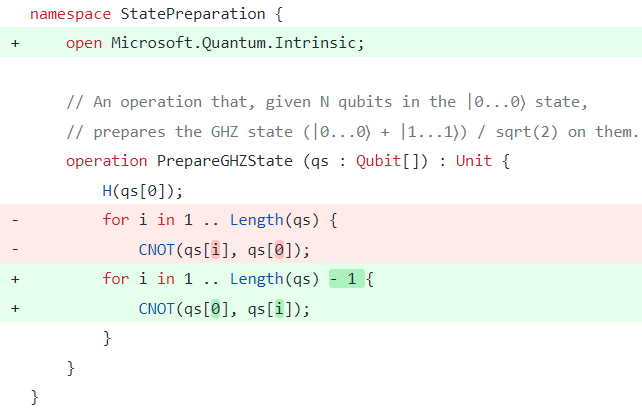}
    \caption{Debugging task that shows three errors in the GHZ state preparation code: a syntax error (missing open statement), a runtime error (incorrect iteration range), and a logical error (swapped control and target qubits in the CNOT gate)}
    \label{fig:debugging-task}
\end{figure}

\subsection{Debugging task example}

Figure~\ref{fig:debugging-task} shows an example of a debugging task, with the errors and the fixes highlighted in red and green, respectively. The task offers the students Q\# code that takes an array of qubits in the $|0\rangle$ state as an input and intends to prepare the Greenberger–Horne–Zeilinger (GHZ) state $\frac1{\sqrt2}(|0...0\rangle + |1...1\rangle)$ on them. The code contains three errors, one of each possible type:

\begin{itemize}
    \item The code fails to compile with the compilation error QS5022 "No identifier with the name "H" exists." This can be fixed by adding the missing \texttt{open} statement for the namespace Microsoft.Quantum.Intrinsic that contains the definitions of the built-in quantum gates.
    \item Upon trying to run the code with this fix applied it throws an exception "Specified argument was out of the range of valid values." This is a runtime error caused by the loop variable \texttt{i} running out of bounds of the array \texttt{qs} on the last iteration. Q\# ranges are inclusive, and Q\# array indexes are 0-based, so on the last iteration the loop attempts to access the array element beyond the last one. This can be fixed by changing the upper bound of the range used in the \texttt{for} loop to \texttt{Length(qs) - 1}.
    \item Finally, with these two errors fixed, the code turns out to prepare the state $\frac1{\sqrt2}(|0...0\rangle + |10...0\rangle)$ instead of the required GHZ state. This is caused by the incorrect order of qubit arguments passed to the CNOT gate: qubits \texttt{qs[0]} and \texttt{qs[i]} should be swapped, with the first one used as the control and the second one as the target.
\end{itemize}

\section{Resource estimation}

The next type of assignments introduces students to another important step of the quantum software development cycle - resource estimation. 
Resource estimation tools provide accurate automated estimates of the resources (typically the number of qubits, the numbers of different types of gates and measurements, and the circuit depth) required to run the given quantum program on a quantum device, even if the program is too large to actually be executed or simulated. 

Resource estimation assignments teach students to use the tools provided by the Microsoft QDK to perform these estimates.
Such assignments can be useful to accompany discussions of different algorithms that solve the same problem or different implementations of the same routine, since they allow the student to compare the resources required by different algorithms or implementations side by side.

\subsection{Resource estimation task example}

A resource estimation task could offer the student two implementations of Shor's integer factorization algorithm using different approaches to performing modular arithmetic in the order finding oracle\footnote{https://github.com/microsoft/Quantum/tree/main/samples/algorithms/integer-factorization}, and ask them to compare the resources required to run these implementations to factor several integers of increasing size.

\section{Noise exploration}

Noise exploration assignments introduce students to the concept of quantum noise and the different types of noise that can occur in quantum systems.
Noise simulator provided by the Microsoft QDK\footnote{https://docs.microsoft.com/azure/quantum/machines/noise-simulator} allows for simulating the behavior of Q\# programs under the influence of various noise models configured using the QuTiP library\footnote{https://qutip.org/}.

The assignments could ask students to implement an algorithm and explore its behavior assuming a specific noise model, or, conversely, identify the noise present in the system based on the histogram of the program outputs produced by running it using the noise simulator.

\begin{figure}
    \centering
    \includegraphics[width=.48\linewidth]{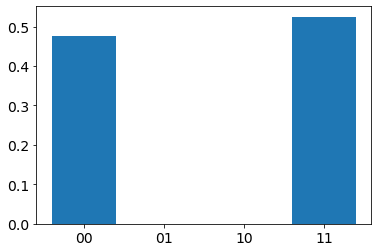}\hfill
    \includegraphics[width=.48\linewidth]{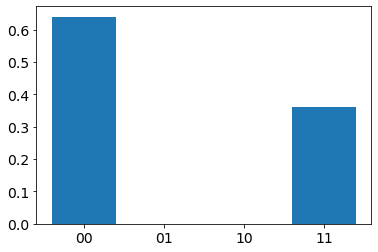}
    \\[\smallskipamount]
    \includegraphics[width=.48\linewidth]{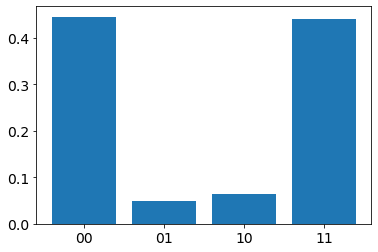}\hfill
    \includegraphics[width=.48\linewidth]{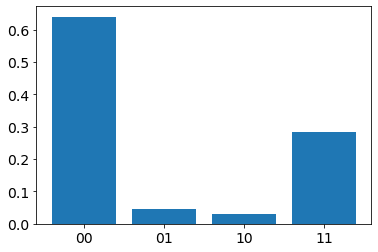}
    \caption{Noise exploration task that asks the student to match the histograms of results yielded by a program that prepares a Bell state and measures both qubits and the noise models used to simulate this program: no noise, amplitude damping noise, bit flip noise, and both noise models at once}
    \label{fig:noise-exploration-task}
\end{figure}

\subsection{Noise exploration task example}

Figure~\ref{fig:noise-exploration-task} shows four histograms of results yielded by the program that prepares a Bell state and measures both qubits, produced with different noise models: 

\begin{itemize}
    \item no noise (perfect simulation)
    \item amplitude damping noise associated with the Hadamard gate: applying a Hadamard gate to the qubit has a 20\% probability of replacing that qubit's state with the $|0\rangle$ state
    \item bit flip noise associated with the CNOT gate: applying a CNOT gate to a pair of qubits has a 10\% probability of additionally flipping the state of the target qubit
    \item both the amplitude damping noise and the bit flip noise described
\end{itemize}

The task is to match the noise models with the histograms produced and to explain the reasoning.

\section{Quantum hardware exploration}

Quantum hardware exploration assignments present an interesting challenge to the instructor. On the one hand, students are always excited to run a program on a real quantum device, so such assignments are attractive for them. On the other hand, quantum devices are currently a scarce resource that requires the student to plan their work on the assignment ahead to make sure they get the results of their jobs back in time to complete the assignment. Besides, a lot of pedagogical value that can be derived from running programs on a quantum device can be delivered by local or cloud simulators, such as the state vector simulator or the noise simulator.

\subsection{Quantum hardware exploration task examples}

Here are several ideas for assignments that focus on quantum hardware exploration via Azure Quantum in a general introductory course, and the concepts they teach:

\begin{itemize}
    \item Explore the concept of noise and the fact that it impacts the results of the computation. The task could suggest running the same simple program (for example, preparing an entangled state, measuring all the qubits, and plotting the histogram of the results) on a noiseless simulator and on a real quantum device and comparing the results.
    \item Explore the size limitations of the programs that can be executed on a quantum device to produce a result that is better than a random sample. The task could suggest implementing an algorithm that solves a specific problem (for example, Grover's search algorithm that identifies a variable assignment satisfying the given SAT formula) and increasing the size of problem solved by it while observing the probability of getting the correct answer.
    \item Explore and compare the limitations of kinds of programs that can run on different devices. In Azure Quantum, these limitations are described as \textit{target profiles}. "No Control Flow" profile allows the user to run only programs that return measurement results directly to the caller, "Basic Measurement Feedback" profile allows the programs to use the measurement results in simple conditional statements, and "Full" profile allows for arbitrary code execution. The task could ask to implement a given algorithm to run on targets with different profiles and to compare the implementations. For example, the teleportation protocol can be implemented straightforwardly on a target with "Basic Measurement Feedback" profile, but the targets with "No Control Flow" profile do not allow the program to apply fixup gates to the resulting state. Instead, running the teleportation protocol on these targets requires implementing the post-selection variant of the protocol that discards the results of teleportation unless both Alice's measurements yielded 0.
\end{itemize}

\section{Conclusions and future work}

Based on our experience in these courses, I conclude that the described software-oriented approach is an effective way to introduce the students to quantum programming. 
The students enjoyed the hands-on assignments and named the Quantum Katas as the component of the course that contributed the most to their learning.
Our teaching assistants in the CSE490Q course expressed their appreciation of the automated  grading tools that reduced their workload considerably\cite{UW}.

I am working on incorporating noise exploration and hardware exploration tasks in the future offerings of the CSYE6305 course, and evaluating their effectiveness in teaching the corresponding topics of the course.

Additionally, the artifacts created for our courses are released to university professors worldwide to inspire them and help them adopt a similar teaching approach\footnote{https://github.com/microsoft/quantum-curriculum-preview}.

\section{Acknowledgements}  

I thank Krysta Svore who invited me to help prepare and deliver the course "CSE490Q: Introduction to Quantum Computing and Quantum Programming in Q\#" at the University of Washington, and Kal Bugrara who invited me to teach the course "CSYE6305: Introduction to Quantum Computing with Applications" at the Northeastern University. I thank the Computer Science and Engineering department at the University of Washington and the College of Engineering at the Northeastern University for enabling these courses. Finally, a special thanks to our students for their valuable feedback.

\bibliographystyle{IEEEtran}
\bibliography{library,other}

\end{document}